\documentclass[conference]{IEEEtran}
\IEEEoverridecommandlockouts

\usepackage{hyperref}
\usepackage{cite}
\usepackage{amsmath,amssymb,amsfonts}
\usepackage{graphicx}
\usepackage{textcomp}
\usepackage{xcolor}
\usepackage{algorithm}
\usepackage{algorithmic}
\usepackage[utf8]{inputenc}
\usepackage{amsmath} 
\usepackage{pgfplots}
\pgfplotsset{compat=1.18}
\usepackage{pgfplotstable}
\usepackage{caption}
\usepackage{adjustbox}
\usepackage{float}

\def\BibTeX{{\rm B\kern-.05em{\sc i\kern-.025em b}\kern-.08em
    T\kern-.1667em\lower.7ex\hbox{E}\kern-.125emX}}

\makeatletter
\newcommand{\linebreakand}{%
  \end{@IEEEauthorhalign}
  \hfill\mbox{}\par
  \mbox{}\hfill\begin{@IEEEauthorhalign}
}
\makeatother

\begin{document}

\title{HashKitty: Distributed Password Analysis}

\author{\IEEEauthorblockN{Pedro Antunes}
\IEEEauthorblockA{
\textit{Polytechnic University of Leiria}\\
Leiria, Portugal \\
2211045@my.ipleiria.pt}

\and

\IEEEauthorblockN{Tomás Santos}
\IEEEauthorblockA{
\textit{Polytechnic University of Leiria}\\
Leiria, Portugal \\
2201762@my.ipleiria.pt}

\linebreakand

\IEEEauthorblockN{Daniel Fuentes}
\IEEEauthorblockA{
\textit{Polytechnic University of Leiria}\\
Leiria, Portugal \\
daniel.fuentes@ipleiria.pt}

\and

\IEEEauthorblockN{Luís Frazão}
\IEEEauthorblockA{
\textit{Polytechnic University of Leiria}\\
Leiria, Portugal \\
luis.frazao@ipleiria.pt}

}

\maketitle

\begin{abstract}
This article documents the HashKitty platform, a distributed solution for password analysis based on the hashcat tool, designed to improve efficiency in both offensive and defensive security operations. The main objectives of this work are to utilise and characterise the hashcat tool, to develop a central platform that connects various computational nodes, to allow the use of nodes with different equipment and manufacturers, to distribute tasks among the nodes through a web platform, and to perform distributed password analysis. The results show that the presented solution achieves the proposed objectives, demonstrating effectiveness in workload distribution and password analysis using different types of nodes based on various operating systems and architectures. The architecture of HashKitty is based on
a scalable and modular distributed architecture, composed of several components such as computational nodes, integration and control software, a web platform that implements our API, and database servers. In order to achieve a fast and organised development process for our application we used multiple frameworks, runtimes and libraries. For the communication between the computational nodes and the other software we made use of websockets so that we have real-time updates between them.
\end{abstract}


\section{Introduction}
In today's world cybersecurity plays an essential role in protecting systems, networks, and data from unauthorised access, theft, and damage. With its mechanisms, cybersecurity, provides protection, risk management and mitigation to ensure the confidentiality, integrity, and availability of data. It is also important to note that as threats become more sophisticated and widespread, it is essential to implement stronger security measures.

A good security exercise that a specialist can use to identify and exploit system vulnerabilities is penetration testing, which aims to simulate an attack, allowing the discovery of weaknesses in a system's defences that attackers could exploit. This group of specialists seeking to explore and identify vulnerabilities are known as Purple Teams. These teams combine the skills of Red and Blue Teams to simulate malicious attacks or penetration testing to identify security vulnerabilities, working together to improve an organisation's overall security posture.

With the growing concern about computer security, the use of cryptography has increased in all aspects of digital life. The increasing need to analyse and enhance security in password usage has become a critical priority for the industry, reflecting the importance of protecting sensitive data against increasingly sophisticated cyber threats. Hashcat is known as one of the most popular hash analysis software available, mainly using graphics cards to process and analyse data, operating both in Open Computing Language (OpenCL), used with Advanced Micro Devices (AMD) graphics cards or Central Processing Units (CPU) in general, and Compute Unified Device Architecture (CUDA), used with Nvidia graphics cards.

This project contributes to the field of cybersecurity and the academic/industry community in general, as this solution is an open-source password analysis tool, available on GitHub \footnote{\url{https://github.com/luisfrazao/hashkitty}} for the general public.

\section{Related Work}
In this section, we compare existing password cracking tools such as John the Ripper (JtR) and Hashcat, as well as distributed password cracking platforms like Hashtopolis and Fitcrack, highlighting the advantages of our platform, called HashKitty.

\subsection{John the Ripper vs Hashcat}
JtR and Hashcat are two of the most widely used password-cracking tools. JtR is versatile and supports various encryption formats, making it suitable for a broad range of password cracking scenarios. However, Hashcat offers superior performance, especially when leveraging GPU acceleration, and supports a broader range of hashing algorithms.

JtR operates in three main modes: Single Crack, Wordlist, and Incremental. The Single Crack mode uses information from UNIX password files to guess passwords. The Wordlist mode utilises a predefined list of words and their variations, like a dictionary attack. The Incremental mode, which is the equivalent of a brute force attack, tries all possible combinations, making it the most comprehensive but also the slowest \cite{harish_john_ripper}.

Hashcat, on the other hand, supports multiple attack modes including Dictionary, Brute Force, Combinator, Hybrid, and Rule-Based attacks. It can leverage the power of GPUs for faster processing, making it a more efficient choice for large-scale password cracking operations. Hashcat requires explicit specification of the hash type but offers a wide range of supported algorithms and optimisation techniques \cite{hashcat}.
 
\begin{figure} [!ht]
    \centering
    \includegraphics[width=0.8\linewidth]{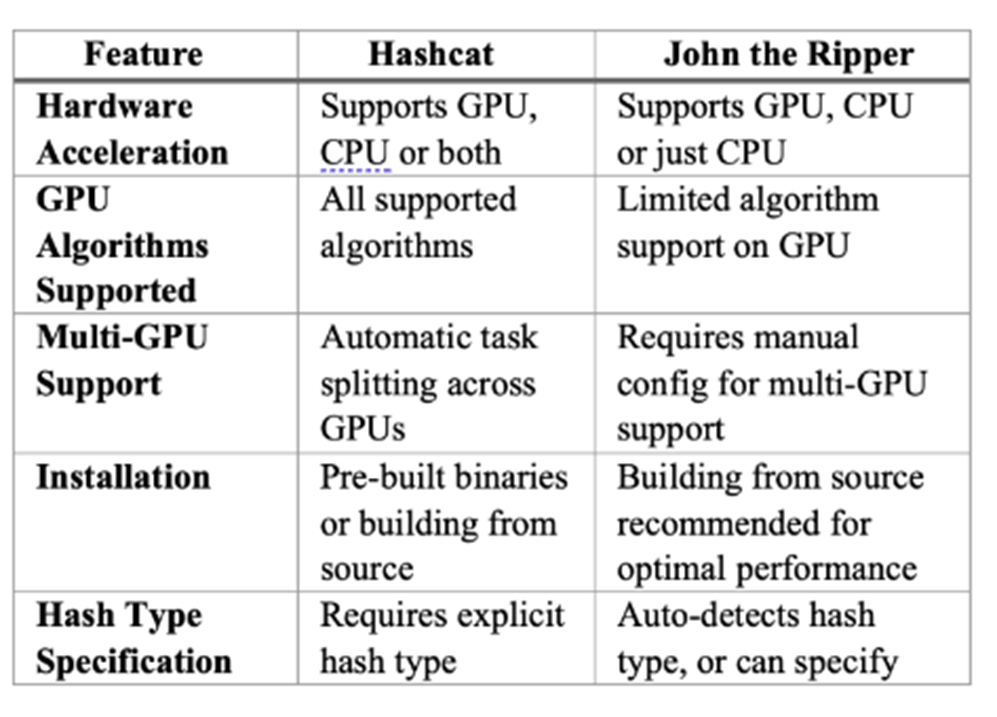}
    \caption{JtR vs Hashcat}
    \label{fig:TableJtrVsHashcat}
\end{figure}

When comparing Hashcat and JtR, Hashcat stands out by supporting a broader range of algorithms on the GPU, enabling more efficient use of graphic processing power. Additionally, Hashcat facilitates multi-GPU setups by automatically splitting tasks across GPUs, whereas JtR requires manual configuration for such support. In terms of installation, Hashcat offers pre-built binaries or the option to build from source, while JtR recommends compiling from source for optimal performance. Furthermore, Hashcat requires explicit specification of the hash type, while JtR can automatically detect or allow specification of the hash type. Overall, Hashcat excels in its flexibility, ease of use, and performance, making it the superior choice for cryptographic and password security tasks. These can be viewed in the Fig.\ref{fig:TableJtrVsHashcat}\cite{JtrvsHashcat}.

\subsection{Hashtopolis and Fitcrack}
Hashtopolis and Fitcrack are platforms that manage distributed password cracking tasks. 

Hashtopolis is a cross-platform client-server tool designed to distribute Hashcat tasks across multiple computers, enhancing the management of password analysis. It features a web interface and API for easy task management, detailed statistics on task progress and hash results, efficient data management for organising wordlists and rules, and automatic updates to ensure the tool remains up-to-date. Built to be portable, robust, and support multiple users, Hashtopolis simplifies and optimises the process of managing Hashcat operations \cite{hashtopolis}.

FitCrack combines BOINC and Hashcat to create a scalable, efficient password cracking system. It distributes tasks among clients, supports centralised management, offers real-time monitoring, and automates processes to improve efficiency and reduce manual intervention \cite{fitcrack}.

In summary, Hashtopolis excels in providing a robust, multi-user platform with comprehensive management features and automatic updates, while FitCrack focuses on scalability and efficiency through its integration with BOINC, real-time monitoring, and automation of processes. Both tools are powerful, but the choice between them depends on specific needs for scalability, ease of management, and automation.

\subsection{HashKitty}
HashKitty employs the hashing capabilities of Hashcat while having a distributed architecture for workload distribution. It supports multiple nodes with different hardware configurations and operating systems, distributing tasks efficiently through a user-friendly web interface.
In addition to the previously mentioned features, Hashkitty also offers support for Single Board Computers (SBCs), a functionality absent in the other two options. This additional feature can be a decisive factor for users who need to integrate and employ SBCs in their password analysis infrastructures, allowing processing at the edge (Edge Computing). The differences in supported features among the tools are summarized in Tab. \ref{tab:featcompare}.

\begin{table}[h!]
\centering
\begin{tabular}{|l|c|c|c|}
\hline
\textbf{Feature} & \textbf{Hashtopolis} & \textbf{Fitcrack} & \textbf{Hashkitty} \\
\hline
Provides User Experience & \textcolor{green!70!black}{Yes} & \textcolor{green!70!black}{Yes} & \textcolor{green!70!black}{Yes} \\ \hline
Support for Multiple Algorithms & \textcolor{green!70!black}{Yes} & \textcolor{green!70!black}{Yes} & \textcolor{green!70!black}{Yes} \\ \hline
Various Attack Types & \textcolor{green!70!black}{Yes} & \textcolor{green!70!black}{Yes} & \textcolor{green!70!black}{Yes} \\ \hline
Work Statistics & \textcolor{green!70!black}{Yes} & \textcolor{green!70!black}{Yes} & \textcolor{green!70!black}{Yes} \\ \hline
Automatic Updates & \textcolor{green!70!black}{Yes} & \textcolor{green!70!black}{Yes} & \textcolor{green!70!black}{Yes} \\ \hline
Multi-Platform System & \textcolor{green!70!black}{Yes} & \textcolor{green!70!black}{Yes} & \textcolor{green!70!black}{Yes} \\ \hline
Multiple Agents & \textcolor{green!70!black}{Yes} & \textcolor{green!70!black}{Yes} & \textcolor{green!70!black}{Yes} \\ \hline
Agent Usage & \textcolor{green!70!black}{Yes} & \textcolor{green!70!black}{Yes} & \textcolor{green!70!black}{Yes} \\ \hline
Work Division on Brute & \textcolor{green!70!black}{Yes} & \textcolor{green!70!black}{Yes} & \textcolor{green!70!black}{Yes} \\ \hline
Distributed Architecture & \textcolor{green!70!black}{Yes} & \textcolor{green!70!black}{Yes} & \textcolor{green!70!black}{Yes} \\ \hline
Distributed Processing & \textcolor{green!70!black}{Yes} & \textcolor{green!70!black}{Yes} & \textcolor{green!70!black}{Yes} \\ \hline
Multiple Jobs Simultaneously & \textcolor{green!70!black}{Yes} & \textcolor{green!70!black}{Yes} & \textcolor{green!70!black}{Yes} \\ \hline
SBCs Support & \textcolor{red}{No} & \textcolor{red}{No} & \textcolor{green!70!black}{Yes} \\
\hline
\end{tabular}
\caption{Feature comparison between Hashtopolis, Fitcrack, and Hashkitty}
\label{tab:featcompare}
\end{table}

Furthermore our implementation divides the work in a different way from other solutions. In Hashtopolis and Fitcrack they use an option in Hashcat called  "keyspaces" that lets them divide the work, while our division is done by the algorithms we will be mentioning in the following chapter \cite{FITPT890}.

\section{Architecture}
The architecture of HashKitty is based on a scalable and modular distributed architecture, composed of several components. This section details the different components, the algorithms for distributing hashes and dictionaries, and the various types of attacks supported by the platform.

\begin{figure} [!ht]
    \centering
    \includegraphics[width=0.9\linewidth]{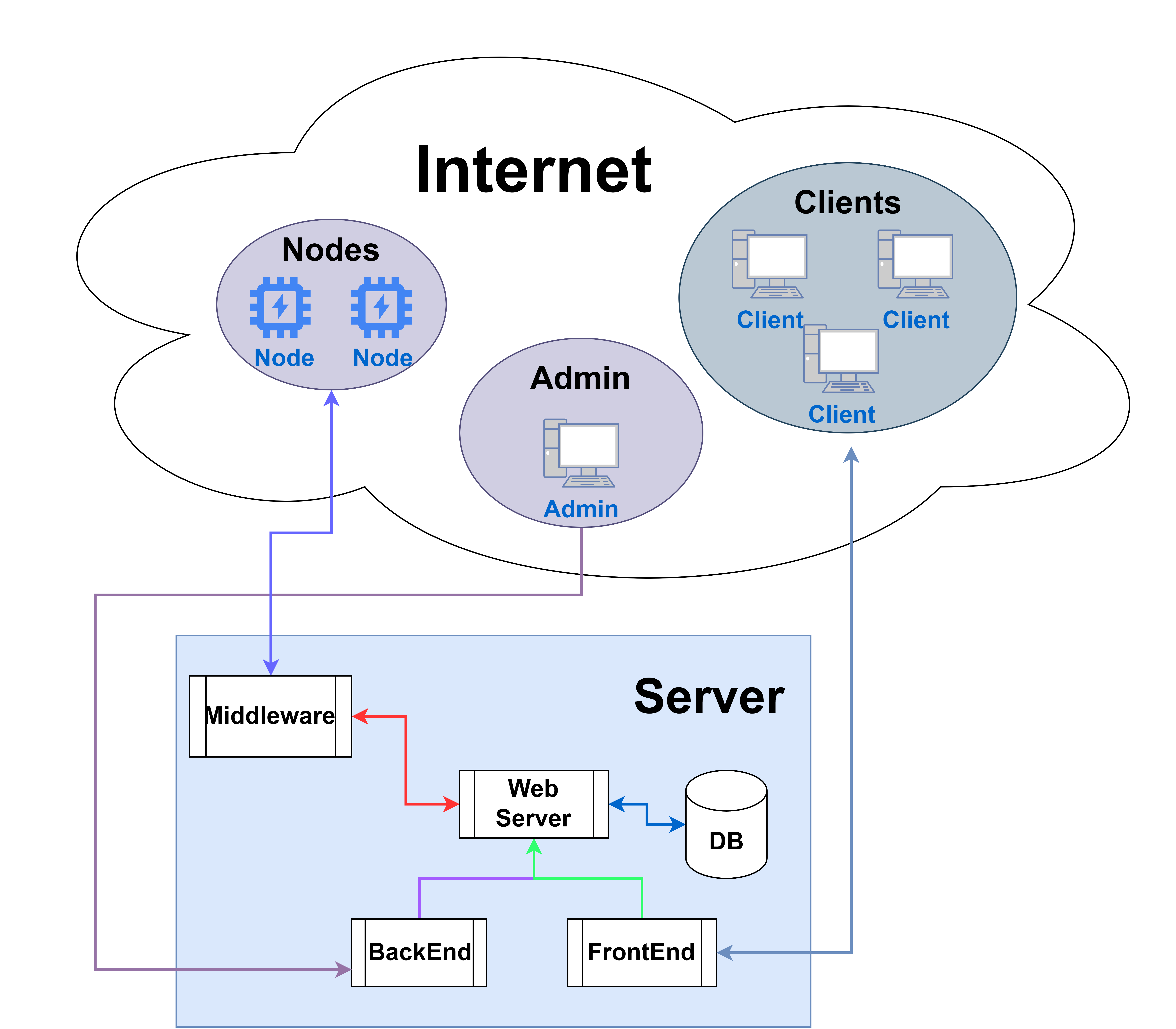}
    \caption{Hashkitty Architecture}
    \label{fig:HashkittyArquitecture}
\end{figure}

\subsection{Components}
HashKitty consists of the following main components:
\begin{itemize}
    \item \textbf{Nodes} - Computational units that perform the actual password cracking tasks. Nodes can be any device with computational power, such as Computers or SBCs;
    \item \textbf{Middleware} - Manages communication between the nodes and the web server, distributing tasks and collecting results. The middleware is responsible for ensuring that tasks are evenly distributed among the nodes, taking into account their computational power;
    \item \textbf{Web Server} - Hosts the web interface and handles HTTP requests. The web server is built using NGINX for its lightweight and scalable nature;
    \item \textbf{Database} - Stores information about tasks, nodes, and users;
    \item \textbf{Backend} - Provides the admin with its interface, making him able to manage nodes, middleware and users, and view application statistics;
    \item \textbf{Frontend} - Provides the client with an interface, allowing them to send work to chosen nodes, view the statistics of each job they have created, and see their own statistics.
\end{itemize}

\subsection{Algorithms for Distribution}

The HashKitty Middleware algorithms distribute hashes and dictionaries across the nodes, taking into account the computational power of each node~\cite{hashkitty_linux}. This ensures that each node receives an amount of hashes or dictionaries corresponding to its computational power, so nodes with low computational power receive less work than those with higher computational power. The distribution algorithms work as follows:

 The Algo.~\ref{alg:distribute_hashes} performs a proportional load distribution of hashes across nodes based on their estimated computational power. This ensures efficient workload balancing and avoids under-utilization of high-performance nodes. Initially, it calculates the total computational power and determines a proportional target count of hashes for each node. After calculating and allocating the primary target counts, it addresses any remaining hashes by distributing them among the most powerful nodes in a round-robin fashion. The algorithm then assigns hashes to nodes according to these target counts, ensuring a balanced workload distribution reflective of each node's capacity. Finally, it returns a distribution map detailing the hashes assigned to each node.

The Algo.~\ref{alg:distribute_dicts} distributes dictionaries among nodes based on their computational power. It begins by defining dictionary sizes and parsing the input list of dictionaries. The total size of selected dictionaries is calculated, and if there are more nodes than dictionaries, the weakest nodes are removed. The total computational power is computed, and initial target sizes for each node are set. Dictionaries are then sorted by size, with the lightest assigned to the weakest node. Remaining dictionaries are distributed based on target sizes and current loads, ensuring each node receives a proportionate workload. Unassigned dictionaries are allocated to nodes with the most remaining capacity, resulting in a balanced distribution of dictionaries across nodes.

\vspace{100px}

\begin{algorithm}
\caption{Distribute Hashes Based on Node Powers}\label{alg:distribute_hashes}
\begin{algorithmic}[1]
\normalsize
\raggedright
\STATE Initialise $\texttt{total\_power}$ as a sum of all values in $\texttt{total\_powers}$
\STATE Initialise $\texttt{distribution}$ as an empty list for each $\texttt{node\_id}$ in $\texttt{total\_powers}$

\FOR{each $\texttt{node\_id}$ in $\texttt{total\_powers}$}
    \STATE Calculate $\texttt{target\_counts[node\_id]}$ as $\max(1, \lfloor \frac{\texttt{total\_powers[node\_id]}}{\texttt{total\_power}} \times \text{length of hashes} \rfloor)$
\ENDFOR

\STATE Initialise $\texttt{allocated\_hashes}$ as a sum of all values in $\texttt{target\_counts}$
\STATE Calculate $\texttt{remaining\_hashes}$ as $\text{length of hashes} - \texttt{allocated\_hashes}$

\STATE Sort nodes by their $\texttt{total\_powers}$ in descending order

\FOR{$i$ from 0 to $\texttt{remaining\_hashes} - 1$}
    \STATE Increment $\texttt{target\_counts[sorted\_nodes}[i \%$ 
    \STATE $\texttt{length of sorted nodes}]]$
\ENDFOR

\STATE Initialise $\texttt{current\_index}$ to 0

\FOR{each $\texttt{node\_id}$ in sorted($\texttt{target\_counts}$ by $\texttt{total\_powers}$ in descending order)}
    \STATE Set $\texttt{count}$ to $\texttt{target\_counts[node\_id]}$
    \STATE Set $\texttt{distribution[node\_id]}$ to $\texttt{hashes}$ from $\texttt{current\_index}$ to $\texttt{current\_index} + \texttt{count}$
    \STATE Increment $\texttt{current\_index}$ by $\texttt{count}$
\ENDFOR

\RETURN $\texttt{distribution}$
\end{algorithmic}
\end{algorithm}

\clearpage

\begin{algorithm}
\caption{Distribute Dictionaries Based on Node Powers}\label{alg:distribute_dicts}
\begin{algorithmic}[2]
\raggedright
\STATE Define \texttt{list\_sizes}
\STATE Split \texttt{dicts} into a list
\STATE Calculate \texttt{total\_size}
\IF{len(\texttt{dicts}) $<$ len(\texttt{total\_powers})}
    \STATE Remove weakest nodes
\ENDIF
\STATE Calculate \texttt{total\_power}
\STATE Initialize \texttt{distribution} for each \texttt{node\_id}
\STATE Calculate \texttt{target\_sizes}
\STATE Sort dictionaries by size
\STATE Assign lightest dictionary to weakest node
\FOR{each \texttt{dict\_name} in \texttt{sorted\_dicts}}
    \STATE Find and assign to node with sufficient target size
\ENDFOR
\FOR{each \texttt{dict\_name} in \texttt{dicts}}
    \IF{not assigned}
        \STATE Assign to node with maximum remaining target size
    \ENDIF
\ENDFOR
\RETURN \texttt{distribution}
\end{algorithmic}
\end{algorithm}


\subsection{Supported Attacks}
HashKitty supports various types of attacks, including:
\begin{itemize}
    \item \textbf{Brute Force} - Tries all possible combinations until the correct password is found. This method is exhaustive but guarantees success if given enough time;
    \item \textbf{Dictionary} - Uses a predefined list of common passwords. This method is faster than brute force but depends on the quality of the dictionary;
    \item \textbf{Rule-Based} - Applies transformation rules to words in a dictionary. This method expands the dictionary with variations of words, increasing the chances of success;
    \item \textbf{Combinator} - Combines 2 chosen dictionaries, combining each word of dictionary 1 to each word of dictionary 2
\end{itemize}

\section{Implementation}
The implementation of HashKitty involves configuring the middleware and agents, and integrating the various components to work seamlessly together.

\subsection{Middleware and Agents}
The middleware is the connection between the nodes and the web server. It distributes the work correctly among the nodes chosen by the client, and it also distributes the hashes and dictionaries uniformly, taking into account the computational power of the nodes. It is implemented in Python due to its extensive library ecosystem, utilising libraries like \textit{websockets}, \textit{requests} and \textit{flask} for making sure we have an efficient communication between the nodes and the web server.

Meanwhile the Agents are nodes that have the python code, Agent.py, currently running. When the Agent starts it tries to establish a connection with the middleware, and after it is successful, it waits for work. Their purpose is to receive a job from the middleware, and execute the proper hashcat command, depending on the hash algorithm and attack mode. Every time one of the hashes it received is discovered, the Agent sends an update to the Middleware, that then sends the discovered hash and its corresponding plain text format, to the API. 

\subsection{Software}
The choice of software is fundamental to ensure the system's compatibility, efficiency, and scalability. Here, we will include a list of platforms, frameworks, libraries, and tools that were used:

\begin{itemize}
    \item \textbf{Hashcat} - Used in the Agents for the analysis of the passwords;
    \item \textbf{Agent.py} – This is the software used by the nodes, programmed in Python, employing libraries such as \textit{websockets} for real-time updates, \textit{detect} so we know what Operating System is being used on the current node, \textit{subprocess} to execute the Hashcat commands, and others;
    \item \textbf{Middleware.py} – This is the software used by the middleware, also programmed in Python, taking advantage of libraries such as \textit{websockets} for real-time updates, \textit{requests} to make \textit{requests} with the web server, \textit{flask} to enable communication between web servers and applications;
    \item \textbf{Nginx} – The chosen web server was Nginx, as it plays an essential role in load balancing and HTTP traffic management, ensuring the scalability and efficiency of our web application;
    \item \textbf{MariaDB} – The MariaDB database system was selected for its free and open-source nature, offering transparency, flexibility, and compatibility with MySQL, making it a reliable and efficient alternative;    
    \item \textbf{Node.js} – The API is developed using Node.js, as the authors believe it provides high performance and the ability to handle intensive asynchronous operations, both of which are essential for system scalability;
    \item \textbf{Vue.js} – Finally, for frontend development, Vue.js was utilised, which facilitates building user interfaces and provides a good user experience.
\end{itemize}

\subsection{Hardware}
To fully achieve the project's objectives, it was imperative to develop a heterogeneous hardware environment, providing the necessary flexibility to accommodate a wide range of technological and operational requirements. This diverse environment allows the demonstration of the performance, scalability, and resilience of our solution, efficiently adapting to the various complexities inherent in the project. For the development of the solution, an environment consisting of the following machines was implemented:

\begin{table}[htbp]
    \centering
    \large
    \begin{tabular}{|l|l|}
    \hline
    \textbf{Node} & \textbf{Specifications} \\ \hline
    Node 1 & Windows 10 - 1 GTX 1660 Super \\ \hline
    Node 2 & Arch Linux - 2 Radeon RX 570 \\ \hline
    Node 3 & Ubuntu 22.04 - 1 RTX 3070 \\ \hline
    Node 4 & Raspberry Pi 5 - ARM Cortex-A76 \\ \hline
    \end{tabular}
    \caption{Nodes Specifications}
    \label{table:hardware_environment}
\end{table}

\section{Results}

In this chapter, we present the testing methodology used to validate our solution, along with the obtained results. We describe the various types of tests conducted, including different operating modes and practical use cases.

\subsection{HashKitty Tests}

Starting with the functional tests of our solution, which supports four types of attacks: Dictionary, Brute Force, Rule-Based, and Combinator. Users can choose between these options before submitting a new task to one or more nodes. Users have the capability to submit hashes in two ways: by uploading a text file, where each line represents a hash, simplifying the process for users with many hashes, or by manually entering the hashes in a text box, useful for users with fewer hashes. These options, among others, can be viewed on the user's dashboard.

After successfully submitting a task, users are redirected to a page where they can view the passwords being analysed in real time. Additionally, users can see real-time statistics of the task and download the analysis as a Comma-Separated Value (CSV) file once completed.

Users can also view global statistics visible on their profile. These statistics include the total number of tasks, including active, completed, and failed tasks. Users can also see the percentage of tasks by mode and algorithm, as well as their activity over time.

Finally, the admin can view global statistics for all users on their dashboard, including information about nodes and completed tasks. Additionally, graphs illustrate the use of different modes and algorithms over time.

\subsection{Benchmarking}

\begin{figure}[H]
    \centering
    \includegraphics[width=0.9\linewidth]{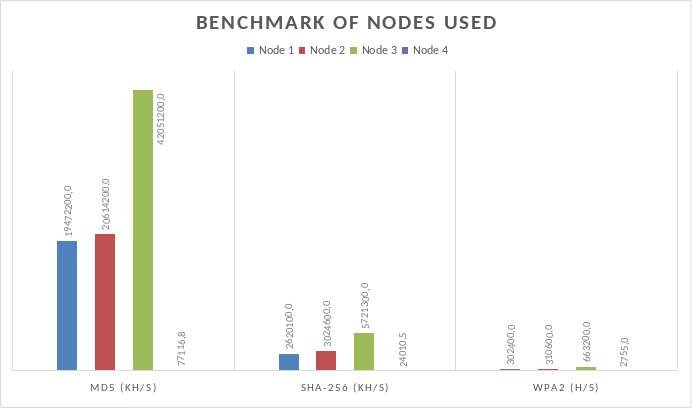}
    \caption{Benchmark of Nodes Used}
    \label{fig:benchmark_nodes}
\end{figure}

This subsection presents a benchmarking analysis for each node in our implementation, demonstrating the performance differences between various devices. Specifically, we showcase the most common use cases in security analysis, including Brute Force, Dictionary, Rule-Based, and Combinator attacks. Our goal with these tests is to evaluate each node's ability to handle the different algorithms chosen for our solution. First, we benchmarked the nodes to determine the number of hashes per second (H/s) each can process, with the results shown in Fig. \ref{fig:benchmark_nodes}.


We used Node 3 as the base to test the different attack types we support (Dictionary, Brute Force, Rule-Based, and Combinator), as it has the highest processing capacity.

To conclude, we will demonstrate the functions developed to calculate the time a node would take based on H/s. It is important to highlight that the values presented below are estimates and not exact numbers, considering the absence of external factors.

Starting with the proposed function for Brute Force, considering there are 95 characters, including  uppercase and lowercase letters, numbers, and special characters, the proposed calculation is as follows, where \( x \) indicates the number of characters in the passwords we want to discover, \( y \) the number of H/s of the node, and \( T \) the time it takes to traverse all possible hashes:
{
\large
\[
T = \frac{95^x}{y}
\]
}

The second proposed function calculates the time to traverse all lines of a dictionary considering the H/s of a node, where \( x \) indicates the number of lines in the dictionary, \( y \) the number of H/s of the node, and \( T \) the time it takes to traverse all lines of the dictionary. \( T \) is not exact, as it does not account for the time taken to load the dictionary, something not considered in this algorithm:
{
\large
\[
T = \frac{x}{y}
\]
}
The third function calculates the time to traverse all lines of a dictionary using rules, where \( x \) indicates the number of lines in the dictionary, \( z \) the number of rules, \( y \) the value of H/s of the node, and \( T \) the time to traverse all permutations:
{
\large
\[
T = \frac{x \times z}{y}
\]
}

The fourth and final proposed function calculates the time to traverse all combinations of two dictionaries, where \( x_1 \) is the number of lines in dictionary 1, \( x_2 \) the number of lines in dictionary 2, \( y \) the value of H/s of the node, and \( T \) the time to traverse all combinations:
{
\large
\[
T = \frac{x_1 \times x_2}{y}
\]
}

By using these functions, a client can estimate the time a specific attack will take by only knowing the available computational power (H/s) from the node, as well as the payload of the attack.

\section{Discussion}
HashKitty benefits from its modular distributed architecture and efficient task distribution algorithms. The use of websockets for real-time communication between nodes and middleware ensures low-latency updates and rapid task reassignment. The system's scalability allows it to handle varying workloads by adding or removing nodes as needed.

One challenge faced during implementation was ensuring compatibility across different operating systems and hardware configurations. This was addressed by using platform-independent technologies and thoroughly testing the system on diverse setups. Future enhancements could include support for additional hash algorithms and integration with other cybersecurity tools to extend the platform's functionality.

\section{Conclusions}
In conclusion, this work  met its objectives by creating a modular and distributed solution utilising various components such as different software tools, frameworks, runtimes and libraries. 
The characterisation of Hashcat  and the analysis of multiple tasks was accomplished through the development of the Agent and Middleware, which handle task management and the distribution of hashes and dictionaries. The solution includes a centralised web platform connecting multiple nodes with different equipment for password analysis.

This project contributes to cybersecurity, academia, and industry by enabling the use of SBCs for security analysis, providing an open-source research tool, and offering an alternative to existing distributed password analysis solutions. Despite achieving the objectives, challenges were faced, such as installing AMD drivers, collecting Hashcat-supported devices, and operating the API implemented in Node.js.

\section{Future Work}
Although the project has achieved the desired results, there are areas that can be improved and new features that can be added. Specifically, future improvements may include: 1) containerization of Agents and Middleware to facilitate their deployment; 2) dynamic addition of dictionaries, where the user would have the possibility to add their own dictionaries to the application; 3) improving statistics; 4) supporting more algorithms and attack types; 5) supporting the use of the hashcatbrain server to optimise the distribution of brute force attacks; 6) allowing a user to cancel tasks during execution; 7) displaying the current average speed of the node in H/s in task statistics; 8) email notifications for completed tasks.

\section*{Acknowledgment}
We would like to express our deep gratitude to everyone who contributed to this work, including our advisers, the Polytechnic University of Leiria, colleagues, friends, and families for their support.

\bibliographystyle{ieeetr}
\bibliography{main}

\end{document}